# Algorithm and Related Application for Smart Wearable Devices to Reduce the Risk of Death and Brain Damage in Diabetic Coma

Vahid Hosseini

Department of Health Sciences and Technology, ETH, Zurich, Switzerland

**Abstract**

Diabetes is an epidemic disease of the 21st century and is growing globally. Although, final diabetes treatments and cure are still on research phase, related complications of diabetes endanger life of diabetic patients. Diabetic coma which happens with extreme high or low blood glucose is one of the risk factor for diabetic patients and if it remains unattended will lead to patient death or permanent brain damage. To reduce the risk of such deaths or damages, a novel algorithm for wearable devices' application, especially for smart watches are proposed. Such application can inform the patient's relatives or emergency centers, if the person falls in coma or irresponsive condition based on readouts from smart watches' sensors including mobility, heart rate and skin moisture. However; such an application is not a final solution to detect all types of coma, but it potentially could save lives of many patients, if widely used among the diabetic patients around the world.

**Introduction**

Diabetic coma is a serious, life-threatening complication of diabetes in which the patient falls into a state of unconsciousness. It constitutes a medical emergency, if left untreated, since it may result in permanent brain damage or death. All insulin dependent diabetic patients are susceptible to the "dead in bed syndrome" a term to describe a person fund dead in bed without clear reason. Especially type 1 diabetic patients, including children are among the high risk group. A study of "causes of death" in children with insulin dependent diabetes in United Kingdom reports 71% of death were caused by diabetes and 10% of those found dead in bed (1). Recent study highlighted type 1 diabetes increases the risk for sudden unexplained death in United States, generating concern that diabetes processes and/or treatments underlie these deaths. In this trial, around 20% (4 out of 19) of sudden death caused by diabetic coma and three of those found dead in bed (2). Both studies indicate that diabetes care still far from its goals despite of advances in insulin delivery and glucose measurement systems. Giving the abovementioned fact, expectation of higher mortality rate of diabetic patients in developing countries is highly likely. Actually, more than 80% of people with diabetes live in low- and middle-income countries (3) and cannot afford implantable glucose sensor for continues blood glucose monitoring which has high initial and maintenance cost (4). Therefore having an adaptable, simple and cheap system to monitor the extreme case of diabetic condition could save many patients' life.

**Algorithm Design and Method**

In this paper, it is describing how to use available technologies including wearable devices such as smart watches to reduce the risk of sudden death in diabetic patients using a novel algorithm. With recent advances in sensors technologies, many smart watches have precise sensing capabilities and connection compatibility to cellular networks. In addition, competitive market of smart watches make them widely available and relatively cheap. Among all sensors on a smart watch, accelerometer, heart rate sensor and humidity sensor were employed to detect the patient physical conditions and were used develop a smart watch application.

*Application output:* The application can inform the patient or relatives by checking the two symptoms of hyper-hypoglycemia if it suspected to diabetic coma. If the patient or relatives dose not respond to the generating alarm, it tries to call out and reach to other relatives or emergency centers.

*Application working logics:* Using smart watch sensors, three body responses were used in algorithm flowchart (**Figure**). The first step is the lack of motion, which is sensed by the accelerometer and shows the status of sleeping, resting or coma. Using suitable data readout, it is even possible to detect tremor from centimeter scale movement and currently are using by many to qualify the users sleep. The program will start automatically after sensing an activity. After an active period, patient immobility will be evaluated. If the application detects a nonmoving condition for few minutes, it starts to read the heart rate data. If there is no heart rate signal the application ends and will restart with another sign of activity. This mostly happens when patient take off the device. This means wearing the device is a compulsory step to activate the application and even the device can remind the patient to wear it if it is around sleeping time. But if the heart rate sensor works, the application will continuously evaluate the heart rate. Tachycardia is one of symptoms of hyper or hypoglycemia and helps to monitor a diabetic patient extreme conditions (5, 6). However; it might happen in case of consuming caffeine, smoking, nightmare or other pathologic conditions. The application would evaluate an increasing heart rate combined to patient immobility as predictive sign of coma or dangerous state and trigger alarm sound. In addition, some smart watches equipped with humidity sensor with the capability of sensing skin humidity. This unique property also can detect skin moisture which is a sign of excessive perspiration in hypoglycemic diabetic patients. Therefore, the application evaluates skin moisture in an immobile patient and unusual increase in skin moisture will be assessed as danger status as well and leads the initiation of an alarm. By triggering a sound alarm, the patient or his/her relatives can deactivate the alarm, monitor his/her blood glucose and/or start other medical intervention. If the application dose not receives any respond, the second alarming system will be activated within a minute. In this step the application will access to cellular network directly and try to contact (call or message) to numbers in its data base including national emergency number in an ordered manner. The two layers alarming system prevents application to make unnecessary calls in case of error or other events.

*functional evaluation using smart watches with heart rate and motion sensors:* To test the application's functionality, Samsung smart watch with heart rate sensor and accelerometer was tried. To emulate a real condition, a healthy person practiced on treadmill to increase the heart rate while fixed the hand on the handle bar to prevent any motion.

**Results**

Using the abovementioned algorithm, an application on Samsung galaxy smart watch and its coordinate's phone was run. The application was triggered alarm after few minutes of practice followed by call to an assigned phone number.

**Conclusion**

The current available technologies to reduce the risk complications after diabetes and diabetic coma are based on continues glucose monitoring. However; such devices have high initial and maintenance cost which is main barrier for diabetic patients to use them in low income families and developing countries. Therefore many diabetic patients are vulnerable to diabetic coma and related complications. Till now, there is no widely accessible and cheap solution to reduce the risk of complications after diabetic coma. By widespread availability of smart watches, a solution to reduce such incidences n case of diabetic coma was proposed. Although, this algorithm and related application cannot guaranty to cover all diabetic coma conditions either because of sensors malfunction or the lack of abovementioned symptoms in patients, it can potentially detect unusual patient conditions. As at the moment there are no other means of helping device in such incidences, we believe if such monitoring system widely used by diabetic patients, could have huge impact on their health and further damages. Having in mind the current statistic of diabetes, in which 382 million of people are living with diabetes in the world, saving life of small portion of patients' lives means thousands of lives every year. Moreover, the widespread availability of smart watch with affordable prices for low income families could help such applications spread easily. The final goal of this paper is to stimulate NGOs, associations and developers and manufacturer to use similar algorithms and implement freely available applications for different wearable smart electronics, as axially monitoring system for diabetes care. Finally the author would like to indicate this algorithm and related application is not a medical device and has to be used as a health monitoring system to complement diabetes monitoring in critical and extreme conditions.

**Acknowledgment:**

The author would like to thanks Alireza Tabasi for developing the example application and verifying its functionality and Professor Micheal Ristow for his encouraging discussion and comment.

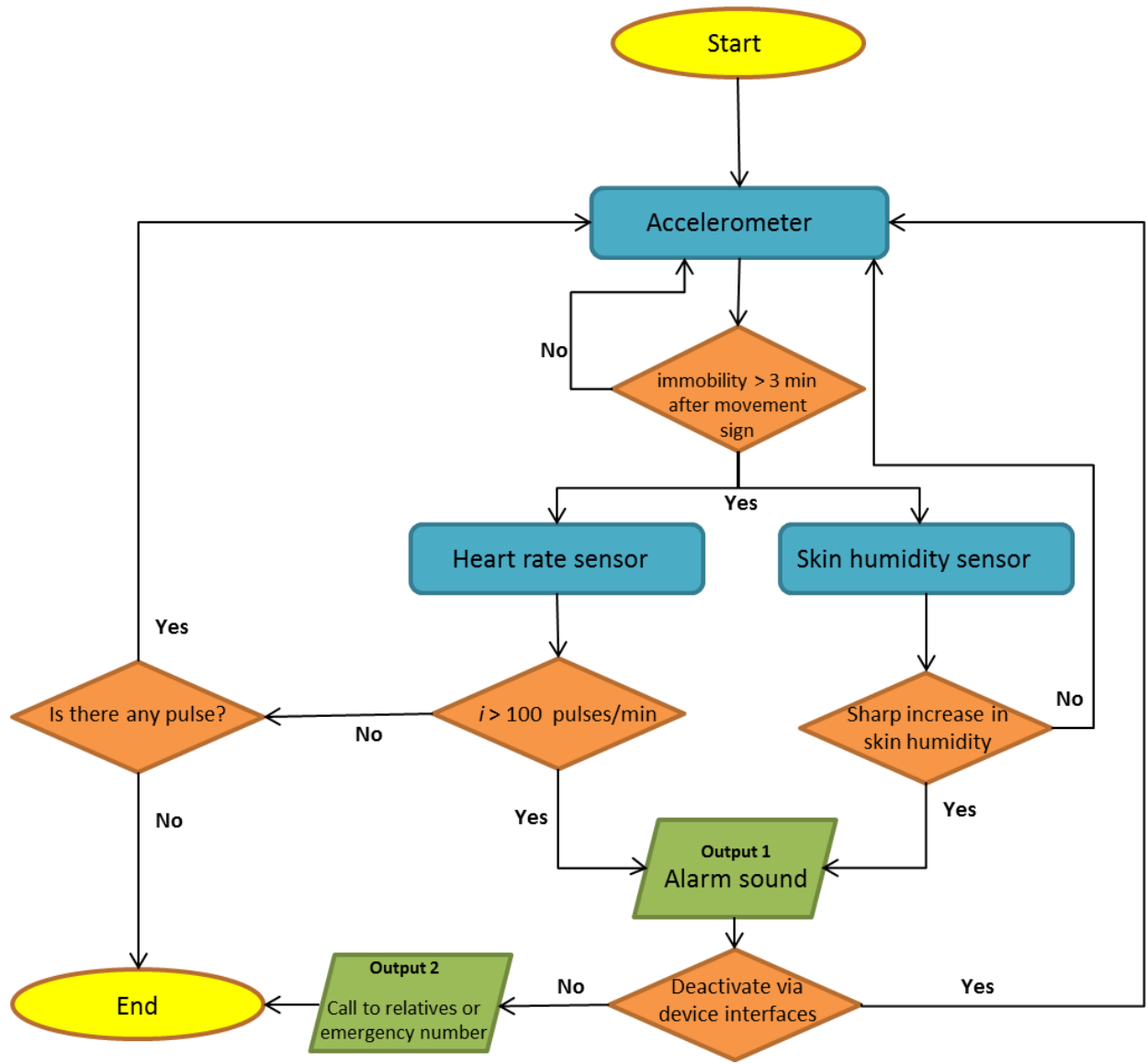

**Figure**: Application's algorithm flowchart and descriptions.